\documentclass{article}
\usepackage{graphicx}

\begin{document}


\newcommand{\BM}[1]{
  \mbox{\boldmath$#1$}
}

\newcommand{\figureTopBot}[1]{
  \begin{figure}[!tb]{\sloppy #1}\end{figure}
}

\newcommand{\figureTop}[1]{
  \begin{figure}[!t]{\sloppy #1}\end{figure}
}

\newcommand{\figureBot}[1]{
  \begin{figure}[!b]{\sloppy #1}\end{figure}
}

\newcommand{\figureWideTop}[1]{
  \begin{figure*}[!t]{\sloppy #1}\end{figure*}
}

\newcommand{\eqAlgn}{
  \!\!&\!\!
}

\newcommand{\comment}[1]{\textbf{ -- #1 -- }}



\date{\small Published as: James F. O'Brien and Jessica K. Hodgins. ``Animating Fracture''. Communications of the ACM, 43(7):68–75, July 2000.}

\title{Animating Fracture}

\author{James F. O'Brien \and Jessica K. Hodgins}

\maketitle


\figureBot{
  \centerline{\includegraphics[width=\columnwidth]{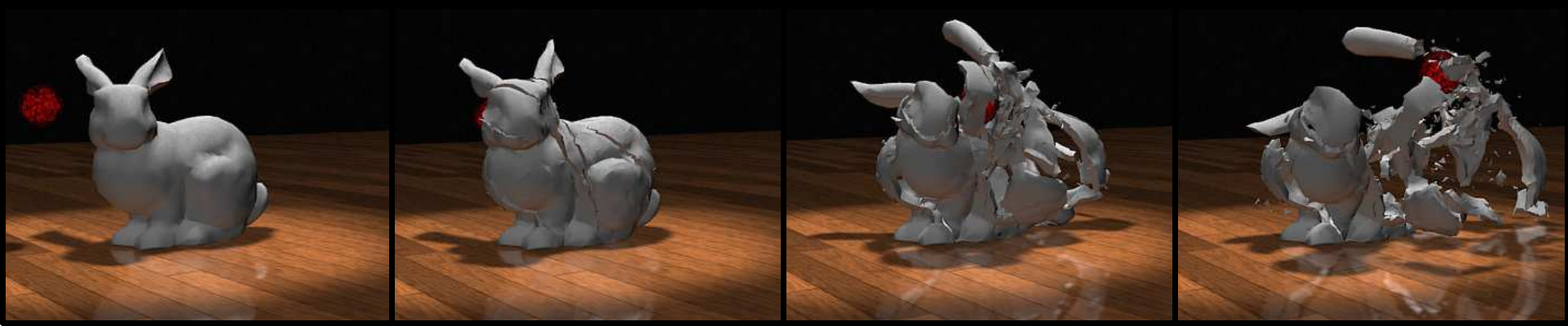}}
  \centerline{(a)}
  \centerline{~~~}  
  \centerline{~~~}  
  \centerline{\includegraphics[width=0.6 \columnwidth]{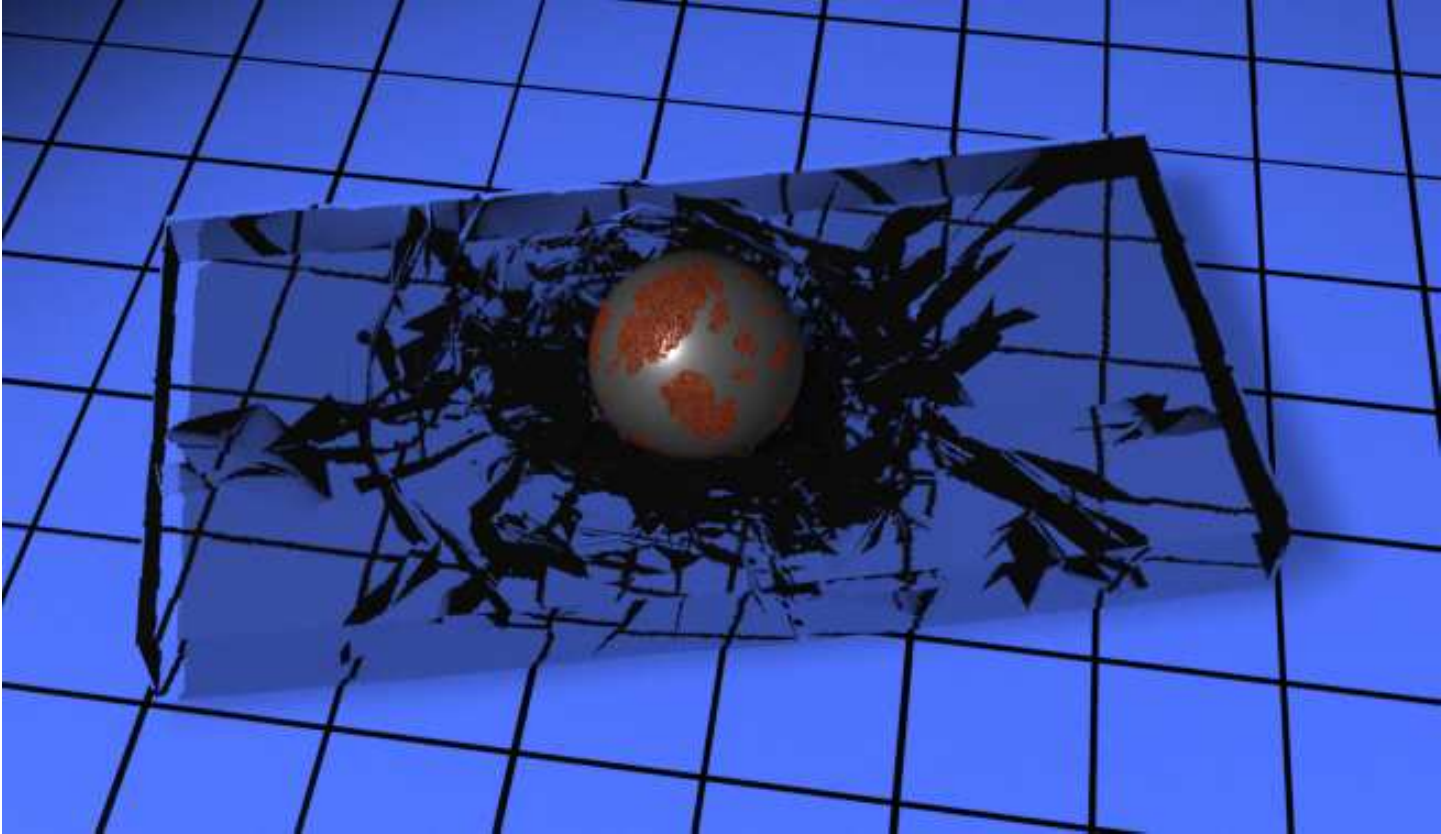}}
  \centerline{(b)}
  \caption{
    These images show the results of using our technique to simulate
    the behavior of \textbf{(a)} a hollow, ceramic bunny as it is
    stuck by a heavy, fast-moving weight, and \textbf{(b)} a slab of
    simulated glass that has been shattered by a heavy weight.
  }\label{fig:BunnyGlassStrip}
}

The task of specifying the motion of even a simple animated object,
like a bouncing ball, is surprisingly difficult.  In part, the task is
difficult because humans are very skilled at observing movement and
quickly detect motion that is unnatural or implausible.  Additionally,
the motion of many objects is complex and specifying their movement
requires generating a great deal of data.  For example, cloth can bend
and twist in a wide variety of ways, and the breaking bunny statue
shown in Figure~\ref{fig:BunnyGlassStrip}.a involves many hundreds of
individual shards.

Three primary techniques are used to generate synthetic motion:
keyframing, motion capture, and procedural methods.  Both keyframing
and motion capture require that the motion be specified by some
external source while procedural methods use an algorithm to
automatically compute original motion.  Many procedural methods are
based on informal heuristics, but a subclass known as physically based
modeling makes use of numerical simulations of physical systems to
generate synthetic motion of virtual objects.  With the introduction
of simulated water in {\it Antz}\,\cite{Robertson:1998:Antz} and
clothing in {\it Stuart Little}\,\cite{Robertson:1999:BBM}, physically
based modeling was clearly demonstrated to be a viable technique for
commercial animation.

Physically based modeling is especially effective for animating
passive objects because they are inanimate objects without an internal
source of energy.  The advantage of using simulation for passive
objects is not surprising, as they tend to have many degrees of
freedom, making keyframing or motion capture difficult.  Furthermore,
while passive objects are often essential to the plot of an animation
and to the appearance or mood of the piece, they are not characters
and do not require the same control over the subtle details of the
motion.  Therefore, simulations in which the motion is controlled only
by initial conditions, physical equations, and material parameters are
often sufficient to produce appealing animations of passive objects.

The computer graphics literature contains many examples of passive
systems that have been successfully modeled with simulation.
Computational fluid dynamics models have been used to animate
splashing water, rising smoke, and explosions in the air.  Footprints
and other patterns left by objects falling on the ground have been
modeled by representing the surface as a height field.  Clothing, hair
and other flexible objects have been approximated with spring and mass
systems or collections of beam elements.

We have developed a simulation technique that uses non-linear finite
element analysis and elastic fracture mechanics to compute physically
plausible motion for three-dimensional, solid objects as they break,
crack, or tear.  When these objects deform beyond their mechanical
limits, the system automatically determines where fractures should
begin and in what directions they should propagate.  The system allows
fractures to propagate in arbitrary directions by dynamically
restructuring the elements of a tetrahedral mesh.  Because cracks are
not limited to the original element boundaries, the objects can form
irregularly shaped shards and edges as they shatter.  The result is
realistic fracture patterns such as the one shown in
Figure~\ref{fig:BunnyGlassStrip}.b.  This paper presents an overview
of the fracture algorithm, the details are presented in our ACM
SIGGRAPH 99 paper\,\cite{OBrien:1999:GMA}.

In the computer graphics literature, two previous techniques have been
developed for modeling dynamic fracture caused by deformations.
In~1988, Terzopoulos and Fleischer introduced a general technique for
animating viscoelastic and plastic
deformations\,\cite{Terzopoulos:1988:MID}.  They modeled certain
fracture effects by severing the connection between two nodes when the
distance between them exceeded a threshold.  They demonstrated this
technique by simulating sheets of paper and cloth that could be torn
apart.  In~1991, Norton and his colleagues broke a model of a china
teapot using a similar technique\,\cite{Norton:1991:AFP}.  Both of
these animation techniques produced good results, especially
considering the computational resources available at the time.  But
both techniques allow fractures only along the boundaries in the
initial mesh structure, creating artifacts in the crack pattern. These
artifacts are particularly noticeable when the discretization follows
a regular pattern, creating an effect similar to the ``jaggies'' that
occur when rasterizing a polygonal edge.

Fracture has been studied more extensively in the mechanics
literature.  Fundamentally, a material fractures when the forces
acting on an atomic level are sufficiently large to overcome the
inter-atomic bonds that hold the material together.  The mechanical
literature contains theories that abstract this small scale
description of fracture to a macroscopic level where it can be used
with a continuum model.  A comprehensive review of this work can be
found in the book by Anderson\,\cite{Anderson:1995:FM} and the survey
article by Nishioka\,\cite{Nishioka:1997:CDF}.

The requirements in graphics and engineering are very different.
Engineering applications require that the simulation predict
real-world behaviors accurately.  In computer animation, simulations
of physical phenomena are tools that allow the animator to realize a
preconceived behavior.  As a result numerical accuracy is less
important and issues relating to visual appearance, ease of use, and
computational efficiency become critical.  Although the techniques
described in this article draw heavily from the fields of fracture
mechanics and finite element analysis, these different requirements
allow simulation techniques for computer animation to make use of
simplifications that would be unacceptable in an engineering context.
Conversely, some of the assumptions used in engineering applications,
such as symmetry, are not acceptable for computer animation.

\section*{Methods}

Our goal is to model fractures created as a material deforms.  First,
we need a model of the material's deformation that provides
information about the magnitude and orientation of the internal
stresses, and whether they are compressive or, conversely, tensile.
Our deformation technique is derived by defining a set of differential
equations that describe the aggregate behavior of the material in a
continuous fashion, and then using a finite element method to
discretize these equations for computer simulation.  This approach is
fairly standard, and many different deformation models can be arrived
at in this fashion.

The continuous model is based on continuum mechanics, and an excellent
introduction to this area can be found in the text by
Fung\,\cite{Fung:1969:FCC}.  The primary assumption in the continuum
approach is that the scale of the effects being modeled is
significantly greater than the scale of the material's composition.
Therefore, the behavior of the molecules, grains, or particles that
compose the material can be modeled as a continuous media.  Although
this assumption is often reasonable for modeling deformations,
macroscopic fractures can be significantly influenced by effects that
occur at small scales where this assumption may not be valid.  For
example microscopic scratches in a glass can concentrate stress
causing fracture in situations where a new, unscratched glass would
not break.  Because we are interested in graphical appearance rather
than rigorous physical correctness, we will assume that a continuum
model is adequate.

There are several different ways to define strain, which measures how
much a material has deformed, and in our work we use the formulation
due to Green\,\cite{Fung:1965:FSM}.  This nonlinear metric only
measures deformation; it is invariant with respect to rigid body
transformations and vanishes when the material is not deformed.  In
addition to the strain tensor, we make use of the strain rate tensor
which measures the rate at which the strain is changing.  It is
defined as the time derivative of strain and exhibits the same
invariant properties.

The strain and strain rate tensors provide the raw information that is
required to compute internal elastic and damping forces, but they do
not take into account the properties of the material.  The stress
tensor combines the basic information from the strain and strain rate
with the material properties and determines forces internal to the
material.  For many materials, a linear relationship between the
components of the stress and strain tensors is adequate.  When
material is isotropic, the linear relationship is parameterized by two
independent variables, $\mu$ and $\lambda$, which are the Lam\'{e}
constants of the material.  The material's rigidity is determined by
the value of $\mu$, and the resistance to changes in volume (dilation)
is controlled by $\lambda$.  Similar parameters relate the strain rate
to damping stress.

When the stress is computed directly from the strain, as described
above, the material behaves elastically meaning that the force exerted
depends only on how much the material has been deformed.  For example,
the force exerted by an ideal spring is completely determined by how
far the spring has been stretched.  In contrast, the stress in a
plastic material depends on how the material has been deformed in the
past.  If a non-ideal spring is stretched too far it will deform
plastically and bend.  Once bent, the force exerted by the spring
depends both on how far it is currently stretched and how it was bent
previously.

Although an elastic relationship between stress and strain is suitable
for the behavior of most materials, plasticity plays an important role
in fracture.  Ductile materials, such as many metals, that tear do so
in large part because of the material's plasticity.  Plastic behavior
can be added by separating the strain into two components, an elastic
and a plastic component.  The plastic strain is subtracted from the
total strain yielding the elastic strain that is then used to compute
the stress according to the elastic stress to strain relationship.

\figureTop{
  \centerline{\includegraphics[width=\columnwidth]{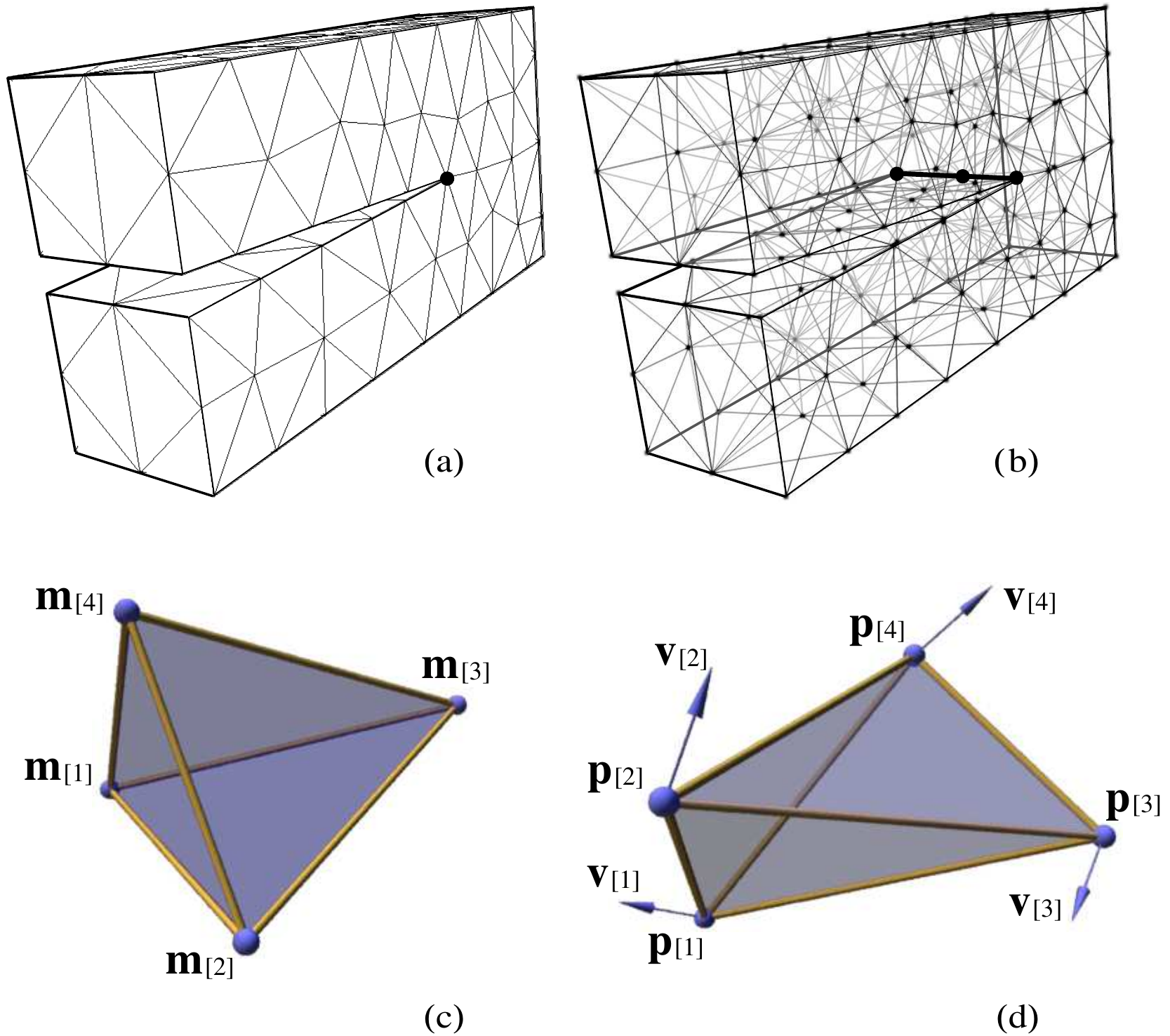}}
  \caption{ 
    Tetrahedral mesh for a simple
    object. In~\textbf{(a)}, only the external faces of the tetrahedra
    are drawn; in \textbf{(b)} the internal structure is shown.
    A tetrahedral element is defined by its four
    nodes.  Each node has~\textbf{(c)} a location in the material
    coordinate system and~\textbf{(d)} a position and velocity in the
    world coordinate system.
  }\label{fig:ElMesh}
}

The definitions used to compute quantities, such as the strain or
strain rate, assume the existence of a function that maps an object
from some reference state to its deformed configuration.  We use a
finite element method to perform this mapping by tesselating the
material into distinct elements, as shown in Figure~\ref{fig:ElMesh}.
Within each element, the material is described locally by a function
with a finite number of parameters that are associated with the nodes
of the element.  Our discretization employs tetrahedral finite
elements with linear polynomial shape functions.  Just as triangles
can be used to approximate any surface, tetrahedra can be used to
approximate arbitrary volumes.  Adjacent elements have nodes in
common, so that the mesh defines a piecewise linear function over the
entire material domain.

Each tetrahedral element is defined by four nodes.  A node has a
position in the material coordinates, a position in the world
coordinates, and a velocity in world coordinates.  (See
Figure~\ref{fig:ElMesh}.c and~\ref{fig:ElMesh}.d.) The continuous
functions that determine strain, strain rate, plastic strain, and
stress within the object are now described in a piecewise linear
fashion in terms of the values at the nodes.  The stress is used to
compute the internal forces acting on the nodes.  These forces
determine the accelerations and the motion of the deformable object is
computed by numerically integrating the entire system forward in time.

This system works with solid tetrahedral volumes rather than with the
polygonal boundary representations created by most modeling packages,
but a number of systems are available for creating tetrahedral meshes
from polygonal boundaries.  The models that we used were generated
either from a CSG (constructive solid geometry) description or a
polygonal boundary representation using NETGEN, a publicly available
mesh generation package\,\cite{Netgen}.

In addition to the forces internal to the material, the system
computes collision forces when two elements intersect or if an element
violates a geometric constraint such as the ground plane.  Determining
which elements intersect can be very expensive, and we use a dynamic
bounding hierarchy scheme with cached traversals to reduce the cost.

To compute the collision forces, we use a penalty method that applies
a force based on the volume of the region formed by the intersection
of two objects.  The overlapping volume is computed by clipping the
faces of each tetrahedron against the other.  A penalty force
proportional to the volume, acting at the center of mass of the
intersecting region is applied to the two tetrahedra.  Provided that
neither tetrahedra is completely contained within the other, this
criteria is quite robust, and the forces computed with this method do
not depend on the object tessellation.  Penalty methods are fast to
compute, but they are often criticized for creating numerically stiff
equations.  However, we have found that the internal elastic forces
are at least as stiff as the penalty forces for the materials in the
examples presented in this article.


Our fracture algorithm is as follows: after every time step, the
system resolves the internal forces acting on all nodes into their
tensile and compressive components.  At each node, the resulting
forces are then used to form a tensor that describes how the internal
forces are acting to separate that node.  If one of the eigenvalues of
this separation tensor is sufficiently large, a fracture plane is
computed.  The resistance of a material to fracture is characterized
by the material's toughness parameter.  When a fracture occurs, the
node is split into two and all elements attached to the node are
divided along the plane with the resulting tetrahedra assigned to one
or the other incarnations of the split node, thus creating a
discontinuity in the material.  The algorithm re-meshes the local area
surrounding the new fracture by splitting elements that intersect the
fracture plane and modifying neighboring elements to ensure that the
mesh remains self-consistent.  This re-meshing preserves the
orientation of the fracture plane and avoids the artifacts seen with
previous methods.

\figureWideTop{
  \centerline{\includegraphics[width=\textwidth]{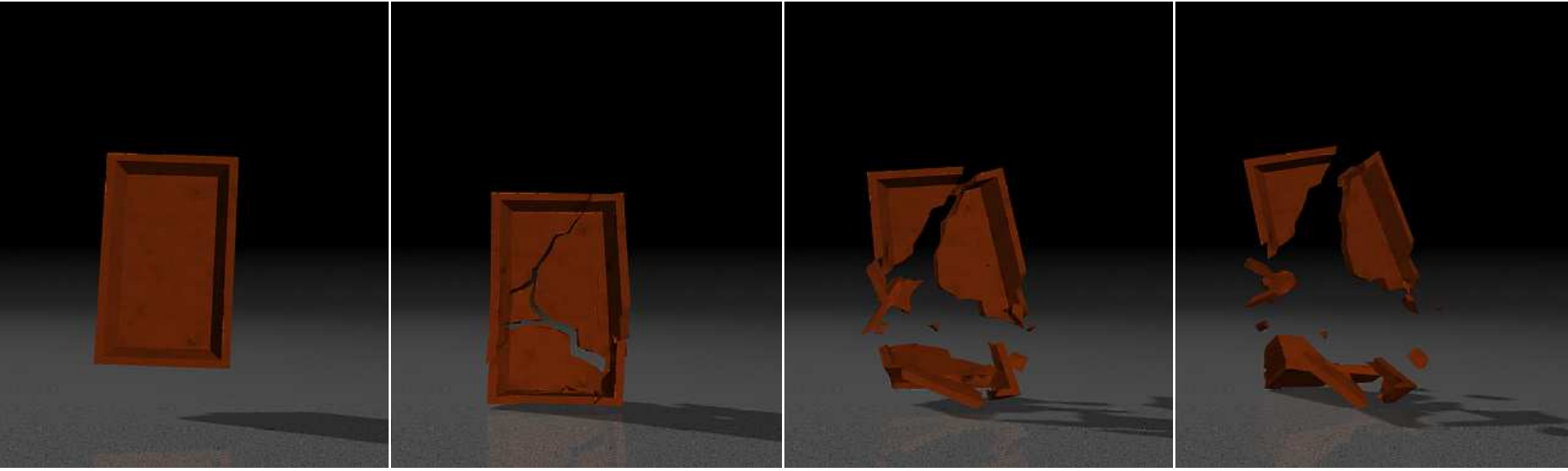}}
  \centerline{\includegraphics[width=\textwidth]{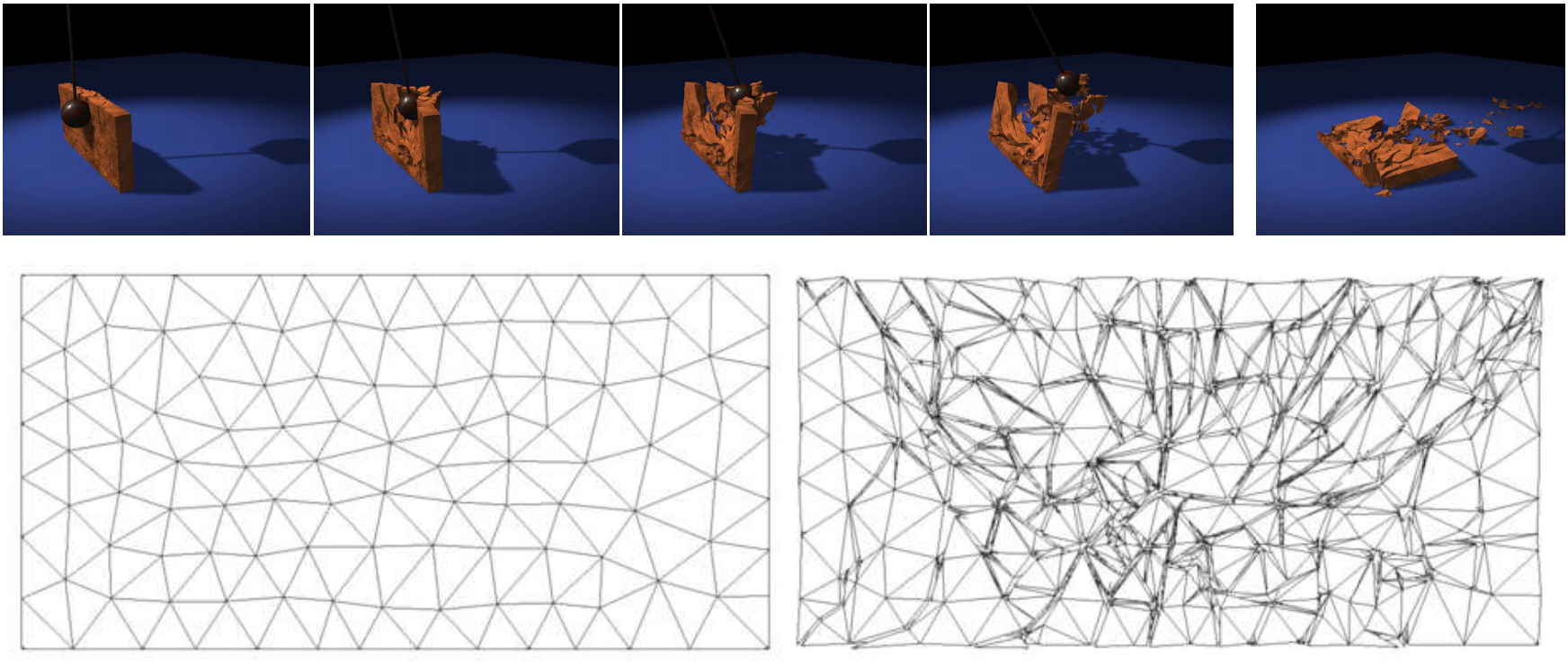}}
  \caption{ 
    \textbf{First Row:}~A terra-cotta flower tray dropped onto a hard floor.  Images are
    spaced $20$\,ms apart.
    \textbf{Second Row:}~An adobe wall struck by a wrecking ball.  
    The bottom edge of the wall is attached to the ground plane.  The
    images are spaced $66.6$\,ms apart.  The rightmost image shows
    the final configurations.  \textbf{Third Row:}~Mesh for the
    adobe wall.  On the left is the facing surface of the initial
    mesh used to generate the wall.  On the right is the reassembled
    mesh after being struck by the wrecking ball.
  }\label{fig:TrayWallAndMesh}
}

\section*{Results and Discussion}

\figureWideTop{
  \centerline{\includegraphics[width=\textwidth]{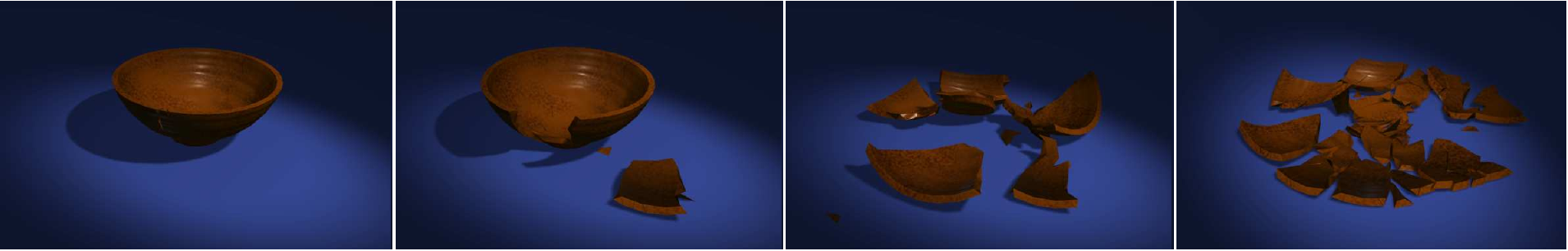}}
  \centerline{\includegraphics[width=\textwidth]{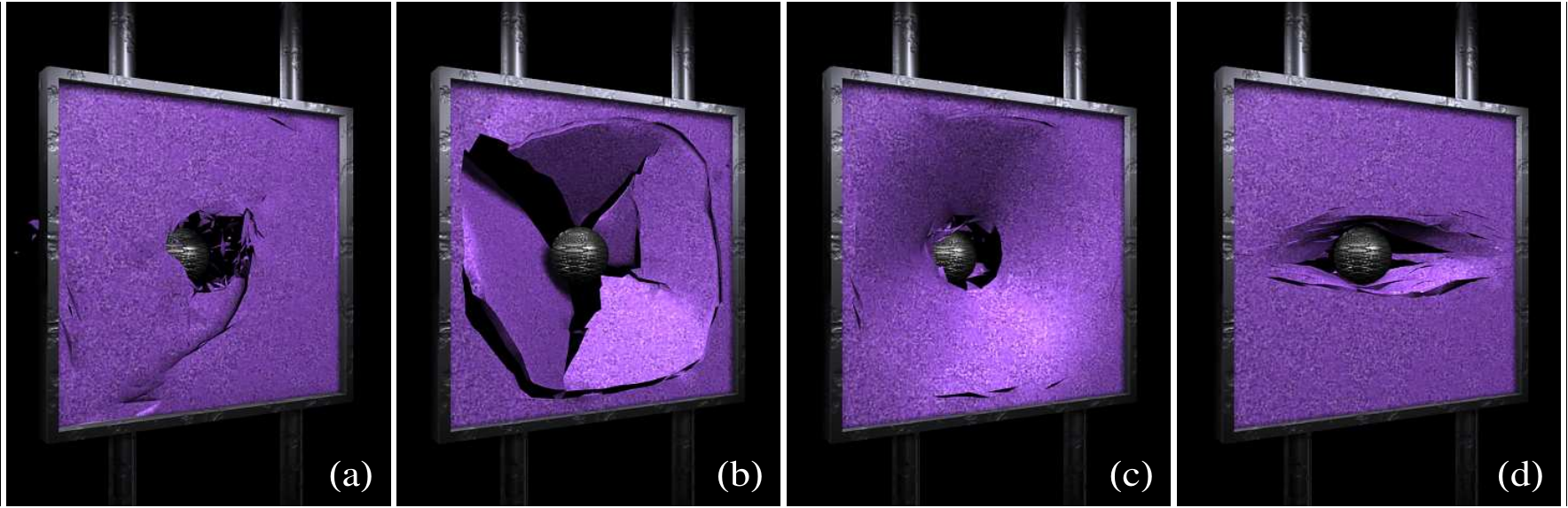}}
  \caption{ 
    \textbf{First Row:}~Bowls with successively lower toughness values
    and otherwise identical material properties.  Each of the bowls
    was dropped from the same height.
    \textbf{Second Row:}~A series of sheets that are struck by a heavy projectile.
    \textbf{(a)}~A stiff, very brittle material.  \textbf{(b)}~A low-density, 
    slightly flexible, but still brittle material.
    \textbf{(c)}~A stiff material that deforms plastically.
    \textbf{(c)}~Another material that yields plastically, but with
    anisotropic toughness.
  }\label{fig:BowlsWindows}
}

We have animated a number of scenes that involve objects breaking to
demonstrate the range of effects that can be generated with this
fracture technique.  Figure~\ref{fig:BunnyGlassStrip}.a shows a
simulation of a hollow, ceramic model of the Stanford Bunny as it is
stuck by a heavy, fast-moving weight.
Figure~\ref{fig:BunnyGlassStrip}.b shows a plate of glass that has had
a heavy weight dropped on it.  The area near the impact has been
crushed into many small fragments.  Farther away, a pattern of radial
cracks has developed.  The first row of
Figure~\ref{fig:TrayWallAndMesh} shows an object that has been dropped
onto a hard surface.

The second and third rows of Figure~\ref{fig:TrayWallAndMesh} show a
wall being struck by a wrecking ball and the mesh used to generate the
wall sequence.  The initial mesh contains only $338$~nodes and
$1109$~elements.  By the end of the sequence, the mesh has grown to
$6892$~nodes and $8275$~elements because additional nodes and elements
are created where fractures occur. A uniform mesh would require many
times this number of nodes and elements to achieve a similar result.

The simulation parameters correspond to physical quantities that can
be adjusted by the user to achieve desired effects.  The first row of
Figure~\ref{fig:BowlsWindows} shows the final frames from four
animations of bowls that were dropped onto a hard surface.  Other than
the toughness of the material, the four simulations are identical.  A
more varied set of materials is shown in the second row of
Figure~\ref{fig:BowlsWindows}.

The physical parameters of the simulation allow a wide range of effects
to be modeled, but selecting the combinations of parameters needed to
produce a desired result is often difficult.  Although many parameters,
such as the material's stiffness, have intuitive meanings when
considered in isolation, the interactions between parameters can be
complex.  Many of the quantities that an animator may wish to
control, for example the violence with which an object shatters, do not
directly correspond to any single parameter.  The solution to this problem
is to develop mappings from intuitive parameters, such as ``shattering
violence,'' to sets of simulation parameters that are automatically
adjusted to produce the desired effect.

Our approach avoids the ``jaggy'' artifacts in the fracture patterns
caused by the underlying mesh, but the results of the simulation are
still influenced by the mesh structure.  The deformation of the
material is constrained by the degrees of freedom in the mesh, which
in turn dictates how the material can fracture.  This limitation will
occur with any discrete system.  The technique also restricts where a
fracture can start by examining only the existing nodes.  As a result,
very coarse mesh sizes might behave in an unintuitive fashion.
However, nodes occur at concavities and sharp features which are the
the locations where a fracture is most likely to begin. Therefore,
with a reasonable grid size, this limitation is not a serious
handicap.

A more serious limitation is related to the speed at which a crack
propagates because the distance that a fracture may travel during a
time step is determined by the size of the existing mesh elements.
The crack may either split an element or not; it cannot travel only a
fraction of the distance across an element.  If the crack's speed
should be significantly greater than the element width divided by the
simulation time step, then a high-stress area will race ahead of the
crack tip, causing spontaneous failures to occur in the material.  We
have developed heuristics for dealing with this situation by
approximating the stresses at the new crack tip and allowing the crack
to advance across multiple elements in a single time step.

A second type of artifact may occur if a crack were being opened
slowly by an applied load on a model with a coarse resolution mesh,
this limitation would lead to a ``button popping'' effect where the
crack would travel across one element, pause until the stress built up
again, and then move across the next element.  Although we have not
observed this phenomena in our examples, developing an algorithm that
allows smooth, slow fracture propagation is an area for future work.

The goal of this work is realistic animation of fracture.  However
assessing subjective quantities such as realism is difficult.  One
possible way to do so is to compare our computer-generated results
with images from the real world.  Figure~\ref{fig:Real} shows
high-speed video footage of a physical bowl as it falls onto its edge
compared to our imitation of the real-world scene.  Although the two
sets of fracture patterns are clearly different, the simulated bowl
has some qualitative similarities to the real one.  Both initially
fail along the leading edge where they strike the ground, and
subsequently develop vertical cracks before breaking into several
large pieces.

\figureWideTop{
  \centerline{\includegraphics[width=\textwidth]{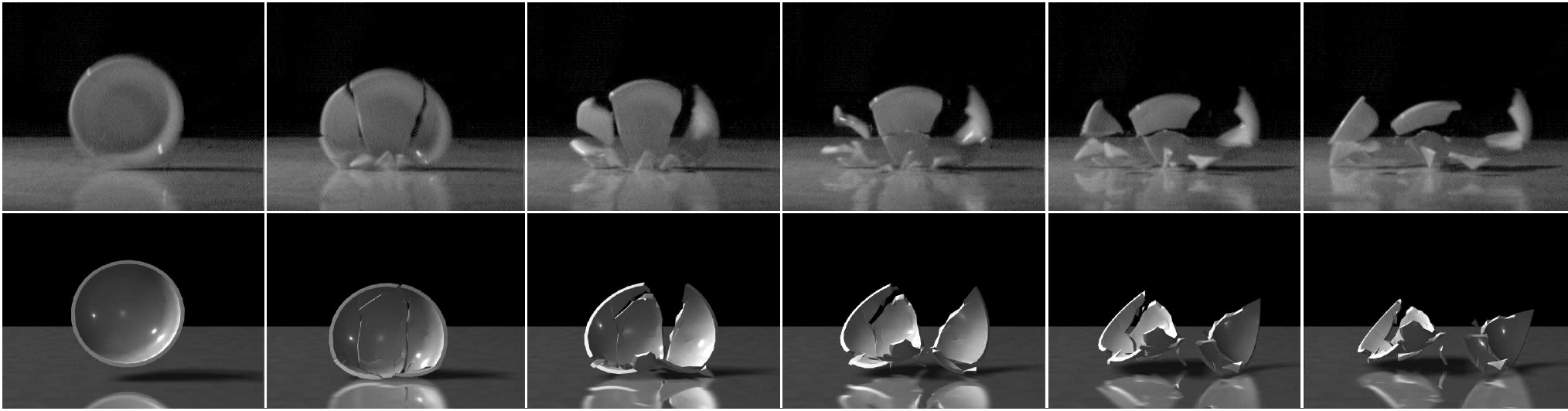}}
  \caption{ 
     Comparison of a real-world event and simulation.  The top row
     shows high-speed video images of a physical ceramic bowl dropped
     from approximately one meter onto a hard surface.  The bottom row
     is the output from a simulation where we attempted to match the
     initial conditions of the physical bowl.  Video images are
     $8$\,ms apart. Simulation images are $13$\,ms apart.
  }\label{fig:Real}
}

\begin{center}
  Additional information about our work on animating fracture and
  animated sequences corresponding to the images in this article can be 
  found at\\
  \texttt{http://www.gvu.gatech.edu/animation/fracture/} 
\end{center}


\bibliographystyle{plain}

\bibliography{fracture}


\clearpage
\vspace{.1in}
\hrule
\vspace{.1in}

\noindent\textbf{James O'Brien} (job@acm.org) is a doctoral candidate
in the College of Computing at the Georgia Institute of Technology
and a member of the Graphics, Visualization \& Usability Center.\\

\noindent\textbf{Jessica Hodgins} (jkh@cc.gatech.edu) is a associate 
professor in the College of Computing and Graphics, Visualization \&
Usability Center at the Georgia Institute of Technology, where she
heads the Animation Lab. \\(See
\texttt{http://www.cc.gatech.edu/gvu/animation}).

\vspace{.1in}
\hrule
\vspace{.1in}

\noindent The authors would like to thank Wayne L. Wooten of Pixar Animation
Studios for lighting, shading, and rendering many of the images in
this article.  We would also like to thank Ari Glezer and Bojan
Vukasinovic of the School Mechanical Engineering at the Georgia
Institute of Technology for their assistance and the use of the
high-speed video equipment.\\

\noindent This project was supported in part by NSF NYI Grant No.~IRI-9457621,
Mitsubishi Electric Research Laboratory, and a Packard Fellowship.
The first author was supported by a fellowship from the Intel
Foundation.


\end{document}